\documentclass[journal=jacsat,manuscript=article]{achemso}
\usepackage[version=3]{mhchem} 

\author{Rachel Yixuan Tan}
\affiliation{Institute of Molecular and Cell Biology (IMCB), Agency for Science, Technology and Research (A*STAR), Singapore 138673}
\author{Rachel Chi Kei Chan}
\affiliation{Institute of Bioengineering and Bioimaging (IBB), Agency for Science, Technology and Research (A*STAR), Singapore 138669}
\author{Whitney Jia Ying Loh}
\affiliation{Institute of Bioengineering and Bioimaging (IBB), Agency for Science, Technology and Research (A*STAR), Singapore 138669}
\author{Kaicheng Liang}
\affiliation{Institute of Molecular and Cell Biology (IMCB), Agency for Science, Technology and Research (A*STAR), Singapore 138673}
\email{liang_kaicheng@imcb.a-star.edu.sg}
\alsoaffiliation{Institute of Microelectronics (IME), Agency for Science, Technology and Research (A*STAR), Singapore 138632}
\alsoaffiliation{Bioinformatics Institute (BII), Agency for Science, Technology and Research (A*STAR), Singapore 138671}
\alsoaffiliation{Lee Kong Chian School of Medicine, Nanyang Technological University (NTU), Singapore 636921}

\title{Miniaturized 2D Scanning Microscopy with a Single 1D Actuation for Multi-Beam Optical Coherence Tomography}

\abbreviations{IR,NMR,UV}

\begin{document}

\begin{abstract}
Miniaturized optical imaging systems typically utilize 2-dimensional (2D) actuators to acquire images over a 2D field of view (FOV). Piezoelectric tubes are most compact, but usually produce sub-millimeter FOVs and are difficult to fabricate at scale, leading to high costs. Planar piezoelectric bending actuators (‘benders’) are capable of much larger actuations and are substantially lower cost, but inadequate for 2D steering. 
We presented a multi-beam fiber scanning platform that generated multi-millimeter 2D scans with a 1D actuator by maximizing the mechanical coupling effect in its orthogonal axis. We further expanded the FOV by demonstrating mosaiced fields driven with spiral and cycloid trajectories, where three optical fibers were optimized to resonate with identical paths in synchronicity. Leveraging optical coherence tomography with a long coherence length laser, we acquired depth-multiplexed images of biological samples at 12.6 µm resolution. This multi-fold improvement in scanning coverage and cost-effectiveness promises to accelerate the advent of piezoelectric optomechanics in compact devices such as endoscopes and headsets, and miniaturized microscopes at point-of-care.
\end{abstract}

\section{Introduction}
Miniaturized imaging devices such as endoscopes are widely used in clinical applications for disease diagnosis and surgical guidance \cite{wallace_minimally_2008, goetz_microscopic_2014}. Commercial endoscopes utilize camera image sensors or fiber optic bundles with 2-dimensional (2D) coverage that are either difficult to miniaturize or suffer from poor image quality \cite{renteria_depixelation_2020}. Alternatively, advanced endoscopes utilize a scanning laser beam over a 2D area for image formation, acquiring one pixel at a time \cite{myaing_fiber-optic_2006}. Such endoscopes may be more compact and integrated with a variety of modalities that provide functional, molecular, or sub-surface structural information. Miniature optical imaging is also of interest in neuroscience as head-mounted systems for real-time monitoring of brain activity \cite{klioutchnikov_three-photon_2022, guan_deep-learning_2022}. 

Miniature laser scanning mechanisms are typically achieved by microelectromechanical systems (MEMS) mirrors/actuators \cite{chen_high-speed_2007, park_forward_2012} or resonant fiber scanners \cite{liu_rapid-scanning_2004, lee_scanning_2010, park_high-speed_2020, rivera_compact_2011, liang_endoscopic_2017}. While these mechanisms enable compact high-speed scanning, high driving voltages and complex fabrication are required, limiting the affordability and accessibility of such devices. High driving voltages exceeding 40-50 V are not ideal due to medical device safety requirements, which forces the use of lower voltages and consequently limits the field of view (FOV) to approximately a few hundred microns width. Sub-millimeter FOVs are impractical in a real-world clinical context and are often insufficient for biological studies of tissue or living models. To increase the FOV, methods such as mosaicing sequential images \cite{hendargo_automated_2013}, wide-field scanning \cite{song_long-range_2016} or parallel imaging with space-division multiplexing \cite{zhou_space-division_2013} were previously reported in non-endoscopic systems. It is challenging to design and manufacture resonant fiber scanners at micro-scales. Fiber scanners traditionally require a millimeter-scale thin-walled piezoelectric tube to generate 2D motion, of which the fabrication is complex, expensive, and difficult to scale. Such devices often suffer from poor fiber alignment and imprecise assembly, affecting the coverage and precision of the scan. Consequently, innovations of such scanner designs has slowed; the seminal work describing the centration of an optical fiber in a piezoelectric tube has remained the most successful implementation since its publication in 2001\cite{seibel_miniature_2001}.

A single dimension of position control should ideally produce a pure 1D motion. However, fiber scanning from 1D actuation has been known to result in nonlinear coupling effects, in which a phenomenon known as `whirling' occurs and distorts the intended scan trajectory. While most groups aimed to eliminate the whirling effect in their fiber scanners to achieve a clean 1D motion \cite{kundrat_high_2011, kaur_scanning_2021}, it has been reported to be possible to produce a stable whirling motion for 2D scanning using a 1D actuator near-resonance \cite{hyer_whirling_1979, haight_stability_2005}. Previous work \cite{huang_resonance-based_2007, wu_realization_2007, wu_two-dimensional_2009} exploited the whirling effect via the addition of rigid structures on the fiber cantilever and reported the generation of Lissajous patterns covering a 2D area using a piezoelectric bender. However, practical imaging demonstrations were limited. This phenomenon has never been fully leveraged for multi-dimensional scanning and data acquisition.

\begin{figure}[htbp]
\centering
\includegraphics[width=\linewidth]{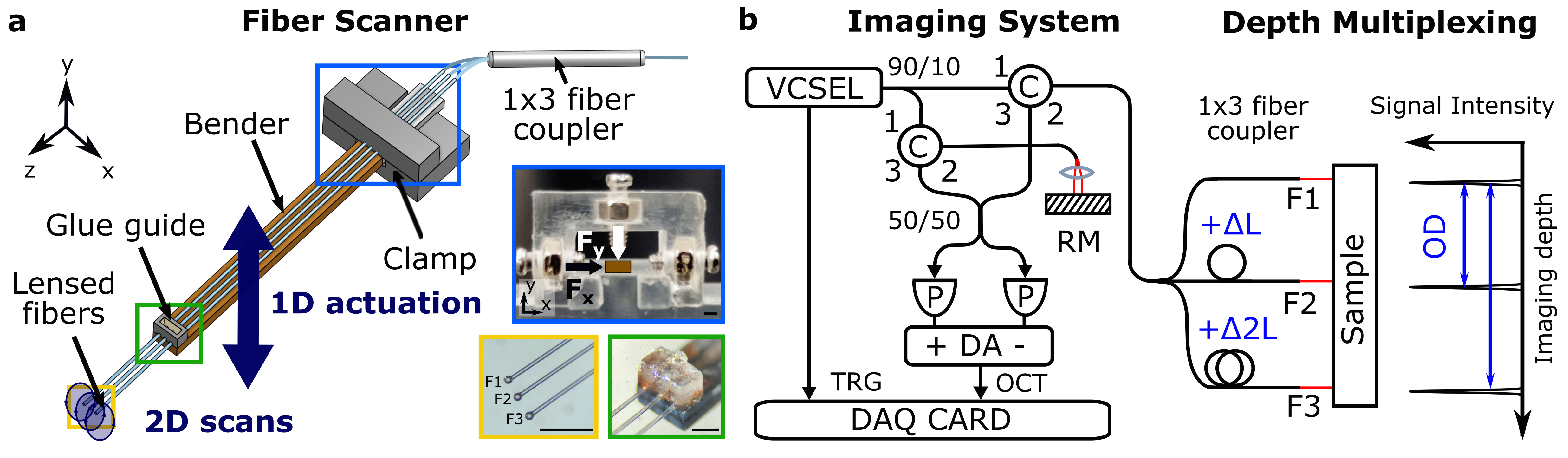}
\caption{Piezoelectric bender fiber scanner and imaging system. \textbf{(a)} Bender fiber scanner generates 2D scan trajectories with a single channel drive. Multiple fibers resonate in tandem to produce mosaiced fields of view. Insets show close up photographs of the lensed fibers (yellow), glue guide (green) and bender clamp (blue). Forces on the bender in the major and minor axes $F_y$ and $F_x$ enable tuning of scan circularity. F: Fiber. \textbf{(b)} Schematic of depth-multiplexed OCT imaging system. Extended fiber lengths +$\Delta L$/$\Delta 2L$ in the sample arm of each imaging beam via a 1x3 fiber coupler produced different optical delays for simultaneous imaging of multiple areas in a single acquisition. VCSEL: Vertical-cavity surface-emitting laser. C: Circulator. RM: Reference mirror. P: Photodetector. DA: Differential amplifier. F: Fiber. DAQ: Data acquisition. $\Delta L$: Extended fiber path length. OD: Optical delay. Scale bars 1 mm, unless specified.}
\label{fig:scanner}
\end{figure}

We present a fiber scanning platform that generated a tunable multi-millimeter 2D scan with a 1D piezoelectric bender, by maximizing the mechanical coupling effect in the axis orthogonal to the main actuation axis (Fig. \ref{fig:scanner}a). This coupling was tuned by adjusting the magnitude of an angled force applied on the bender via orthogonally mounted screws. Two schemes were employed to produce a substantially expanded imaging coverage. In the first approach, a set of resonantly spiral scanning fibers produced multiple FOVs that were mosaiced to form a larger image. Second, the multiple fibers were circularly scanned over a linearly moving sample, effectively generating a cycloid scanning trajectory \cite{liang_cycloid_2018} for large $\sim$ 2 mm wide and arbitrarily long strip FOVs. Depth-multiplexed volumetric imaging (Fig. \ref{fig:scanner}b) with \textit{en face} views was also demonstrated on a wide range of biological samples. Our piezoelectric bender fiber scanning platform is customizable to different scanning fiber schemes and imaging modalities, enabling rich multi-dimensional information to be obtained in a single acquisition.

\section{Methods}
\subsection{Scan mechanism and characterization}
Near-circular laser scanning was an important optimization of the mechanical coupling phenomenon, where scan displacement in the minor axis (coupled axis) nearly equalled that of the major axis (actuation axis). By modulating a single voltage drive, 2D areal coverage could be achieved. An angled force comprising of a preset force in the minor axis $F_x$ and a tunable force in the major axis $F_y$ on the bender allowed a tunable scan from a linear to elliptical to circular scan. To achieve near-identical scans across all fibers, the fibers were driven near-resonance where displacements in the major axis were reduced from maximum to a magnitude similar to that in the minor axis. This relaxed manufacturing requirements for the resonant frequency of each fiber, enabling repeatable and similarly shaped scans despite manufacturing tolerances between each fiber. 

The circularity of a scan ($x/y$ ratio) was defined as the ratio of displacement in minor to major axis of the scan, where circularity of a perfectly circular scan was 1 and a linear scan was 0. To assess the scan shape and size, the scan trajectory of an illuminated fiber tip was captured by a camera and the maximum scan displacements in minor and major axes were measured using ImageJ. The clamping force $F_y$ was measured with a paper-thin force sensing resistor (FSR) connected in a voltage divider circuit and then placed in between the bender and the top clamp.

The bender was driven by a single channel sinusoidal drive of 500 Hz (TD250, Piezodrive) and was controlled by a National Instruments card synchronized with custom software. The circular scan was exploited for 2D areal scanning via two schemes (Fig. \ref{fig:custom}c). The first used a conventional ramping driving voltage amplitude to produce circles of rapidly increasing size, producing a spiral trajectory. The second used a constant driving voltage amplitude to produce circular scans, generating a cycloidal trajectory as the sample was linearly translated using a motor. The effective length of the mosaiced multi-strip FOV could be arbitrarily long (limited only by data acquisition memory). 

The bender was actuated at 500 Hz for characterizing scan circularity at different voltage drives and clamping forces $F_y$. The bender was actuated at 37.5 V for generating Lissajous scan patterns at different clamping forces $F_x$ and driving frequencies (167 Hz, 250 Hz, 500 Hz). The bender was actuated at 15 V for characterizing fibers' frequency responses from 440 Hz to 530 Hz. The clamping forces $F_y$ and $F_x$ were optimized for maximum circularity at 500 Hz. 

\subsection{Scanner assembly}
\begin{figure}[htbp]
\centering
\includegraphics[width=\linewidth]{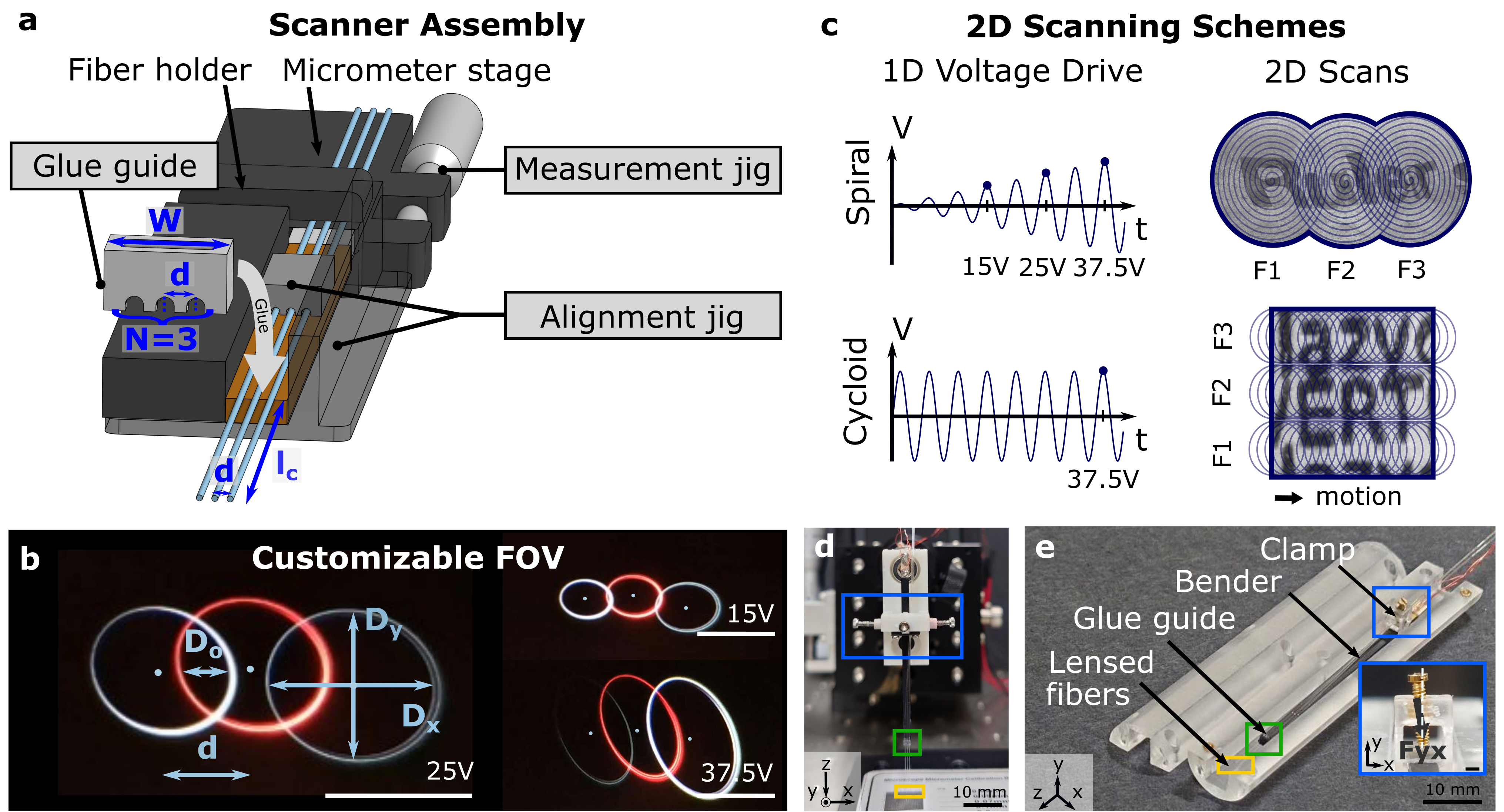}
\caption{Assembly, customizability and implementation of the bender fiber scanner. \textbf{(a)} Scanner assembly setup enabling precise measurement of fiber cantilever length $l_c$ and simultaneous alignment and attachment of multiple fibers on bender. The glue guide can be customized for a different bender width $W$, fiber spacing $d$ and number of fibers $N$. \textbf{(b)} FOV outlines from illuminated fiber tips of the scanner at 25 V, 15 V and 37.5 V. FOVs can be designed given a desired scan overlap between adjacent fibers $D_o$ and scan range in y and x axes $D_y$ and $D_x$, using tradeoffs described in Fig. \ref{fig:charac}a. \textbf{(c)} 2D scanning generated by driving the scanner with a single channel amplitude-modulated (for spiral scanning) or a constant amplitude (for cycloid scanning) sinusoidal waveform. F: Fiber. \textbf{(d)} Bench-top setup of the scanner used in imaging studies. \textbf{(e)} Handheld implementation of the scanner, consisting of a miniaturized clamp (blue) with a single screw mounted at 5 degrees, lensed fibers (yellow) and glue guide (green). Scale bars 1 mm, unless specified.}
\label{fig:custom}
\end{figure}

Three optical fibers were attached on a piezoelectric bender (BA4902, Piezodrive) using a scanner assembly setup consisting of jigs for measurement and alignment, and a glue guide template with fiber spacing $d$ of 0.6 mm (Fig. \ref{fig:custom}a). Given a fixed fiber spacing $d$, the FOVs of scanning fibers could be designed to overlap for image mosaicing or be separated for a maximized FOV using different voltage drives (Fig. \ref{fig:custom}b). The fibers and bender were first carefully positioned in their desired arrangement, parallel to each other, using an alignment jig. Each fiber cantilever length $l_c$ of 14.1 mm was then precisely measured with a micrometer measurement jig to produce a measured resonant frequency of $\sim$ 480 Hz for imaging near-resonance at 500 Hz, such that all three fibers had a near-equal resonant frequency when vibrated in tandem. After which, the glue guide was filled with epoxy and placed over the fibers at the edge of the bender to attach the fiber cantilevers on the bender. The entire assembly process was performed under a stereoscope to ensure a high level of precision and repeatability. 

A ball lens was fabricated at the tip of each fiber, where a coreless fiber (FG125LA, Thorlabs) of splice distance 0.5 mm was first spliced to a single mode fiber (SMF-28 Ultra, Corning). A ball lens of curvature radius 85 $\mu m$ was then created using a Specialty fiber Fusion Splicer (FSM-100P+, Fujikura) \cite{wu_ultrathin_2022}. The spot size and working distance of the ball lens were measured to be 12.6 $\mu m$ (full-width at half-maximum) and 0.35 mm by coupling a laser beam into the fiber and positioning the ball lens in front of a beam profiler such that a minimum focused spot size was observed.

The scanner assembly was mounted in a bench-top clamping structure that consisted of M1.4 screw clamps in orthogonal axes (Fig. \ref{fig:scanner}a). The clamping mechanism enabled precise control and optimization of mechanical coupling from the major axis (actuation axis) to minor axis (coupled axis), achieving reliable and tunable 2D scans. A handheld or endoscopic implementation (Fig. \ref{fig:custom}e) was also constructed to demonstrate potential imaging applications in a compact package. The miniaturized clamping structure consisted of a M1 top screw mounted at a small angle of 5 degrees, where the single angled screw was used to deliver the required clamping forces.

\subsection{Spiral scanning image reconstruction}

\begin{figure}[htbp]
\centering
\includegraphics[width=0.7\linewidth]{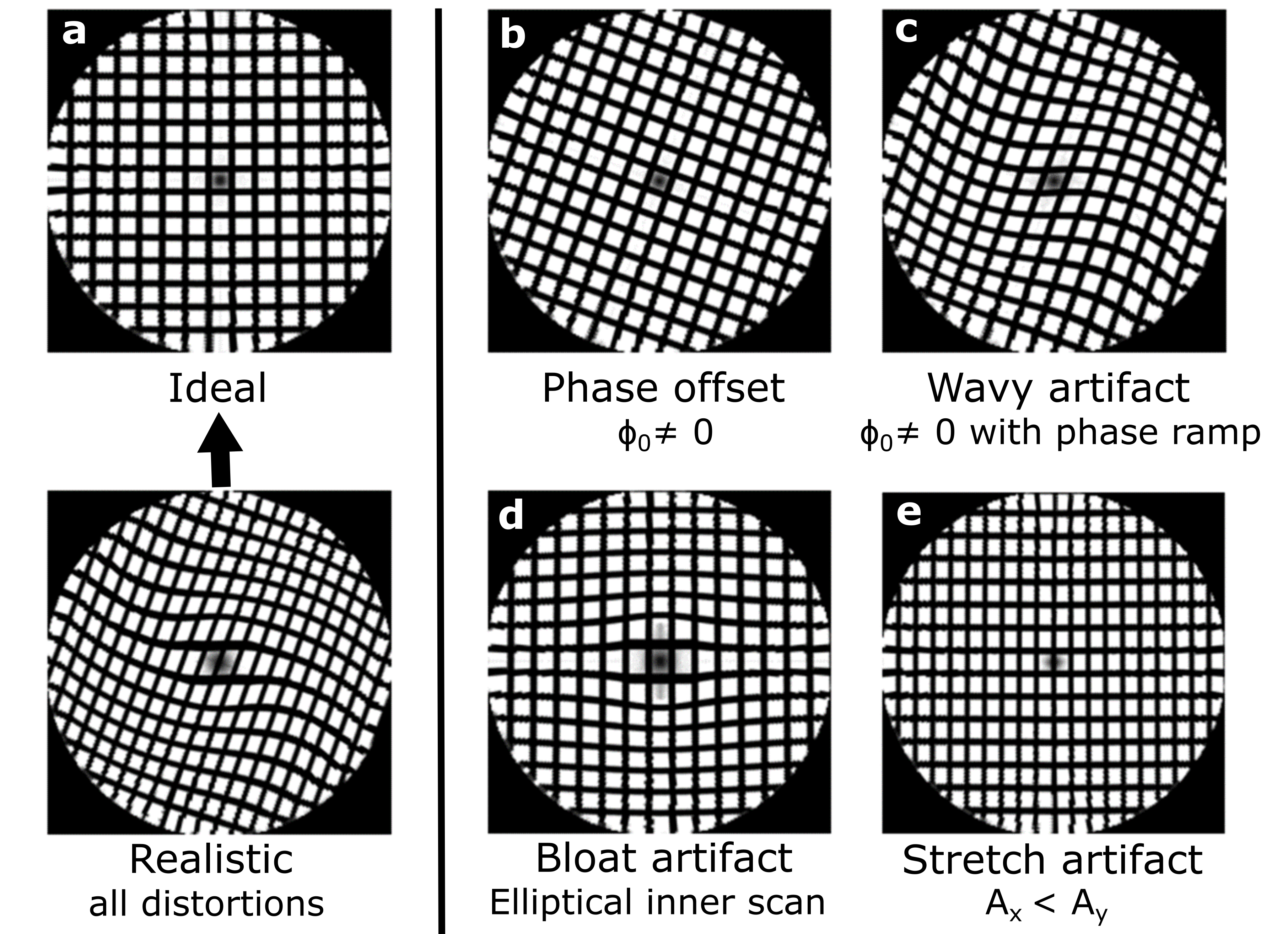}
\caption{Simulated scan artifacts corrected in image reconstruction. \textbf{(a)} Comparison of an ideal image reconstructed from an ideal spiral trajectory and an image with realistic artifacts reconstructed from a distorted spiral trajectory. Distorted images resulting from \textbf{(b)} phase offset $\phi_{0}$, \textbf{(c)} wavy artifact, \textbf{(d)} bloat artifact, and \textbf{(e)} stretch artifact where amplitude of coupled axis $A_x$ is smaller than amplitude of actuation axis $A_y$. }
\label{fig:recon}
\end{figure}

Image data was acquired while the fiber tip traced an approximately spiral trajectory, hence the data was mapped to Cartesian coordinates using a lookup table based on the parametric equations defining a spiral geometry:
\begin{align*}
x &= A_x(t)sin(2\pi f t + \phi_0) \\
y &= A_y(t)sin(2\pi f t + \phi_0 + \phi(t))
\end{align*}
where $A_x(t)=A_y(t)=At/t_0$ (ramp) and $\phi(t)=\pi/2$ for a standard circular spiral, and $\phi_0$ is a phase offset used in distortion correction (described as `Wavy artifact' below). Previous studies of 2D resonant fiber scanners used position-sensing detectors (PSD) to measure the actual motion paths of the fiber \cite{lee_scanning_2010}, in order to capture nonlinear motion effects that would produce image artifacts if not precisely modeled by the lookup table. The scanner's motion could be adequately corrected with a few simple modifications to the standard spiral equations. A set of simulated image reconstructions show an artifact-free grid alongside artifacts that occur in arbitrary combinations in real-world scans (Fig. \ref{fig:recon}a).

\begin{enumerate}
\item \textit{Wavy artifact} (Fig. \ref{fig:recon}c) is produced by a combination of (1) an incorrect $\phi_0$ which can be understood as the angular position (or phase) at which the fiber begins its motion, and (2) a slow-varying $\phi(t)$ phase relationship between the two axes (rather than simply a constant $\pi/2$) leading to rotating, enlarging elliptical scans. In the special case of $\phi(t)=\pi/2$, change in $\phi_0$ produces a pure rotation with no distortion (Fig. \ref{fig:recon}b). $\phi_0$ is stable and does not drift significantly over the course of an imaging session, but may change when mechanical forces on the actuator are adjusted. In the presence of varying $\phi(t)$, an incorrect $\phi_0$ exacerbates distortion. The phase relationship $\phi(t)$ can be modeled as an increasing (or decreasing) linear ramp that eventually reaches $\pi/2$, implying a final circular shape:
\begin{align*}
\phi(t)=
\begin{cases}
  \frac{\pi/2-\phi_1}{\beta_1 t_0} t +\phi_1 & \text{if $t<\beta_1 t_0$} \\
  \pi/2 & \text{if $\beta_1 t_0<t<t_0$}
\end{cases}
\end{align*}
where $\beta_1$ can be interpreted as the fraction of the spiral during which the scan shape is elliptical and can be visually estimated by the diameter of the wavy artifact, and $\phi_1$ is an estimate of the initial elliptical phase.
\item \textit{Bloat artifact} (Fig. \ref{fig:recon}d) occurs in the central area of the field of view and is due to the scan being initially elliptical and eventually becoming circular. It was observed that the coupled axis tended to ramp slower than the actuation axis. This was modeled as follows:
\begin{align*}
A_x(t) &= At/t_0 \\
A_y(t) &= 
\begin{cases}
 \alpha A t/t_0 & \text{if $t<\beta_2 t_0$} \\
\beta_2 \alpha A + \frac{(1-\beta_2 \alpha)A}{t_0-\beta_2 t_0}(t-\beta_2 t_0) & \text{if $\beta_2 t_0<t<t_0$}
\end{cases}
\end{align*}
which more intuitively, the coupled axis is formed by two linear ramp segments joined at $t=\beta_2 t_0$, where the first segment is a ramp with amplitude scaled down by factor $\alpha$ and the second segment catches up to the same peak amplitude $A$. $\beta_2$ can be interpreted as the ratio of first segment to the entire segment and can be visually estimated by the diameter of the bloat artifact, while $\alpha$ can be interpreted as an `ellipticity factor' for the inner scans and is approximately the scale of feature distortion that is created by the bloat artifact.
\item \textit{Stretch artifact} (Fig. \ref{fig:recon}e) occurs when the coupled axis amplitude is consistently lower than the main actuation axis i.e., even the outer scans of the spiral are elliptical. This can be simply modeled by $A_y=\gamma A_x$. 
\end{enumerate}

A real-world scan may be a composite of all of the above described artifacts (Fig. \ref{fig:recon}a). Adequate reconstruction (Fig. \ref{fig:cali}b) may be achieved computationally by optimizing $\phi_0$, $\phi_1$, $\beta_1$, $\beta_2$, $\alpha$, and $\gamma$, without the need for position-sensitive sensor measurements. In this work, these values were obtained by trial and error, but could conceivably be numerically optimized. Higher order (nonlinear) phase or amplitude relationships may be necessary for more precise distortion correction.

\subsection{Cycloid scanning image reconstruction}
Cycloid scanning involved non-modulating circles of fixed diameter, hence reconstruction required fewer parameters than the spiral. The data was mapped to Cartesian coordinates based on parametric equations describing a cycloid trajectory traced by a moving circular scan:
\begin{align*}
x&=Asin(2\pi f t+\phi_0) \\
y&=Acos(2\pi f t+\phi_0) + vt
\end{align*}

where $v$ is the speed of the linear motion. Similar to the spiral scan, $\phi_0$ is the angular position (or phase) at which the fiber begins its motion within the first circle of the trajectory. If left uncorrected, image distortion can be substantial. In practical image reconstruction, the data acquired in time may be circular shifted forward or backward for correction, which has the equivalent effect of shifting the start position of the beam along the trajectory. An inherent characteristic of the cycloid is that the first half of later circles rescan areas already covered by the second half of earlier circles. When reconstructing images incorporating rescanned portions, blur is inevitably introduced, likely because rescans are almost never identical in content to the initial scans and thus do not overlap exactly \cite{liang_cycloid_2018}. It is recommended to use only half of each circular scan for reconstruction. In this way, cycloid scans may produce two `twin' images of the same FOV, where the images have a brief time delay (the time needed for a circle to travel half its own diameter) between them. This could have applications in recovering image content from impulse-like motion perturbations during imaging.

\subsection{Image acquisition and depth-multiplexed OCT system}
The imaging samples in this study were taken on the bench-top setup for optimal performance. The OCT system used a commercial swept-source microelectromechanical system vertical-cavity surface-emitting laser (MEMS-VCSEL) with a center wavelength of 1300 nm (SL132121, Thorlabs). The VCSEL source had a 200 kHz axial scan rate, 8 mm imaging range in air and $\sim$ 100 nm bandwidth. We optimized the axial resolution to $\sim$ 13.8  $\mu$m in air (9.9 $\mu$m in tissue) over the entire depth range, which could be further optimized by dispersion matching and compensation but it was not a primary objective of the present study. The three imaging beams were depth-multiplexed by introducing extended fiber lengths +$\Delta L$/$\Delta 2L$ in the sample arm of each imaging beam via the 1x3 fiber coupler, producing optical delays in the OCT acquisition (Fig. \ref{fig:scanner}b). Enabled by the long coherence length of the VCSEL source, multiple adjacent areas were imaged in a single acquisition over the entire imaging range of 8 mm in air and displayed at different depths in the OCT cross-sectional image. Volumes from each imaging beam could be mosaiced to form large FOVs. The imaging range was realized using a high-speed digitizer (ATS9371, AlazarTech) with 1 GHz bandwidth and 1 GS/s sampling rate, using the optical clock signal provided by the laser. The record length of each axial scan trigger was 1792 samples. Data acquisition and live trajectory-mapped image previews were produced by a custom software written in Python.

Each spiral volume consisted of 400 x 800 axial scans. The number of axial scan samples per circle (400) was determined by the OCT system axial scan rate divided by the driving frequency of the scanner. The number of circles per volume (800) could be arbitrarily set by the ramping time of the circular amplitude from zero to maximum, such that a longer time would permit the repetition of more circular scans (i.e., more radial sampling) within a single volume. Hence the sampling in the circular and radial directions were decoupled. The spiral was undersampled (below Nyquist) in the circular direction due to limited axial scan rate at the design resolution, and oversampled (much over Nyquist) in the radial direction to partly compensate for circular undersampling and to enable more nearest neighbor averaging when mapping the trajectory to a Cartesian grid. Each cycloid volume consisted of 400 x 2000 axial scans, where the number of circles per volume (2000) could also be arbitrarily large, enabling either oversampling or long scan lengths. 

\subsection{Imaging sample preparation}
Calibration targets comprised a printed grid of 100 $\mu$m pitch, a printed word `Ruler', and printed sentences of 0.5 mm and 1 mm font sizes. The targets were prints on a clear plastic substrate overlapped on a sheet of white paper. 
A broad range of biological samples was prepared for imaging, including a finger, pig stomach tissue, an ant, and spheroids. Pig stomach was purchased from a local market and its inner lining was cleaned with phosphate-buffered saline. The stomach antrum was lightly stretched and pinned down by needles to flatten the tissue for imaging. Human breast adenocarcinoma epithelial MCF-7 cell line was purchased from the American Type Culture Collection (ATCC). Formation of spheroids was induced by seeding MCF-7 cells at 2750 cells/well for 700 $\mu$m spheroids in 96-well F-bottomed ultra-low attachment plates with shaking for 2-3 days.

\section{Results}
\subsection{Scan performance optimization}
\begin{figure}[htbp]
\centering
\includegraphics[width=\linewidth]{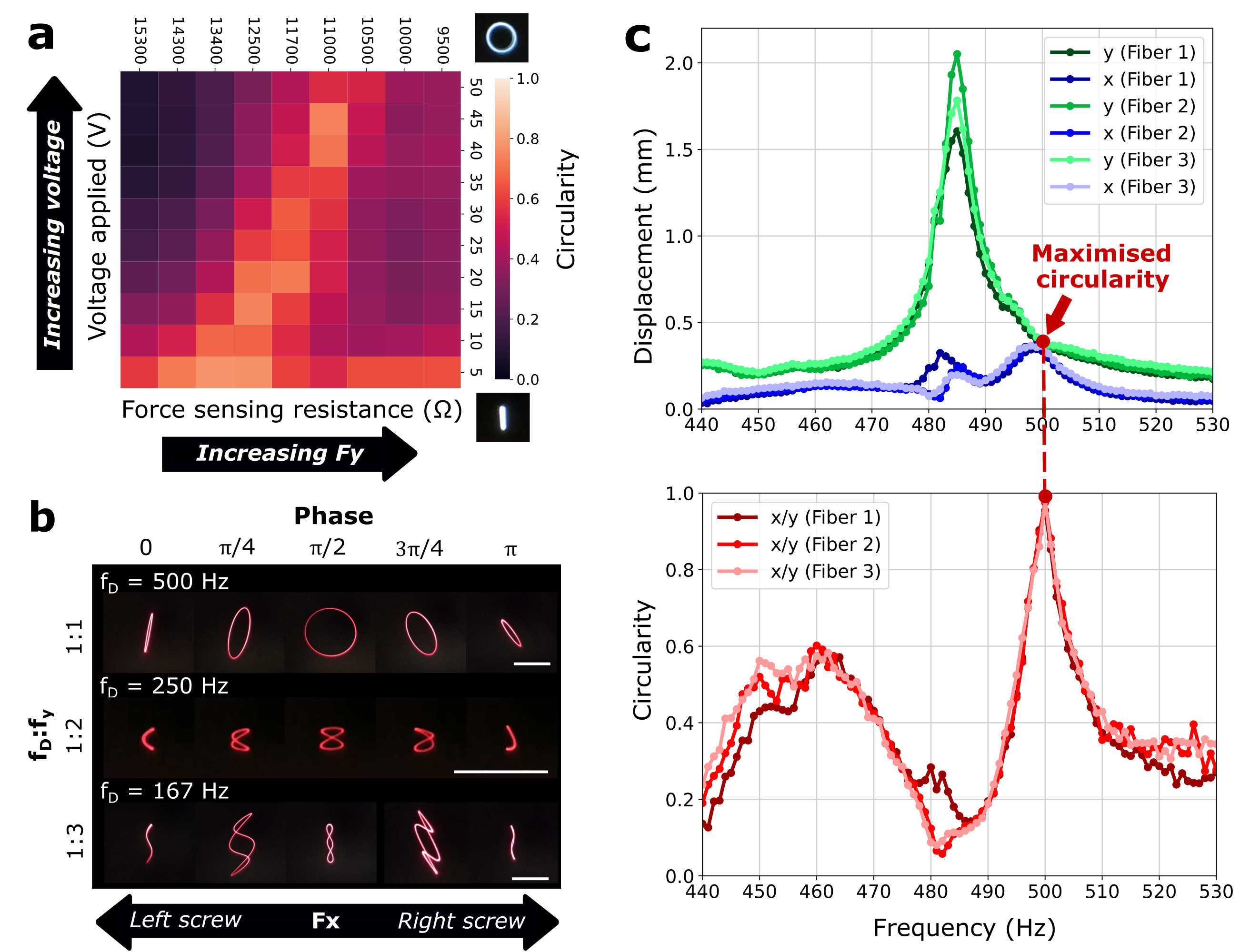}
\caption{Scan characterization. \textbf{(a)} Heatmap showing the relationship of scan circularity with voltage drive and clamping force in the major axis $F_y$, measured in terms of force sensing resistance. An increasing voltage generally resulted in an increasing scan size, whereby a ramping voltage was used to generate a spiral scan trajectory. All measurements were taken at a preset force in the minor axis $F_x$. The bender was actuated at 500 Hz. \textbf{(b)} Lissajous scan patterns at different ratios of driving frequency $f_D$ to designed resonant frequency of the scanning fiber $f_y$, and at apparent phases between the two axes. The phase was controlled by adjusting the force applied in the minor axis $F_x$ in either direction. The bender was actuated at 37.5 V. Scale bars 1 mm. \textbf{(c)} Frequency response plot in the major ($y$) and minor ($x$) axis of three scanning fibers, and a corresponding circularity plot. The circularity ($x/y$) is a ratio of the scan displacement in the minor to major axis. Maximum circularity was tuned to 500 Hz. The bender was actuated at 15 V.}
\label{fig:charac}
\end{figure}

Detailed characterization of the clamping force in the major axis $F_y$ and scanning fiber resonances were performed. The heatmap (Fig. \ref{fig:charac}a) showed the relationship of scan circularity (i.e., degree of minor axis coupling) with increasing voltage drive and clamping force $F_y$. All measurements were taken at a preset force applied in the minor axis $F_x$. Higher voltages generally produced larger scans, in which a ramping voltage drive was used to spirally scan an area for image formation. 

At higher voltages, displacements in the major axis dominated the trajectory, where a larger clamping force $F_y$ was required to maintain the scan circularity. The clamping force $F_y$ had to be sufficiently large to impart a force on the bender, and sufficiently low to allow the bender to vibrate freely for mechanical coupling. There exists an operating regime for near-circular scans over a broad range of voltages and clamping forces, $F_y$ and $F_x$. The eventual peak voltage selected for 2D scanning was determined by several design parameters such as clamping forces $F_y$ and $F_x$, fiber spacing $d$, desired FOV size $D_y$ and $D_x$, and scan overlap $D_o$.

The frequency responses of each fiber (Fig. \ref{fig:charac}c) were similar, which was an important manufacturing outcome facilitated by the scanner assembly setup and glue guide template (Fig. \ref{fig:custom}a). In addition to the frequency response plot, a circularity plot was generated. The circularity plot showed how the peak resonance in the major axis was at a circularity minimum and thus was not a useful operating point. The scans were further tuned by adjusting the clamping force $F_y$ to optimize circularity at the preferred operating point of 500Hz. Given the broad operating regime discussed in Fig. \ref{fig:charac}a, the frequency of the circularity maxima could be tuned using $F_y$ at a given peak voltage and clamping force $F_x$. As an extension of the circular scan, Lissajous patterns (Fig. \ref{fig:charac}b) at different driving frequencies $f_D$ and clamping forces $F_x$ were generated, demonstrating different scanning patterns that could potentially be used in image formation. These scan patterns were not utilized for imaging but were presented for completeness. 

\subsection{Multi-beam image mosaicing}
\begin{figure}[htbp]
\centering
\includegraphics[width=\linewidth]{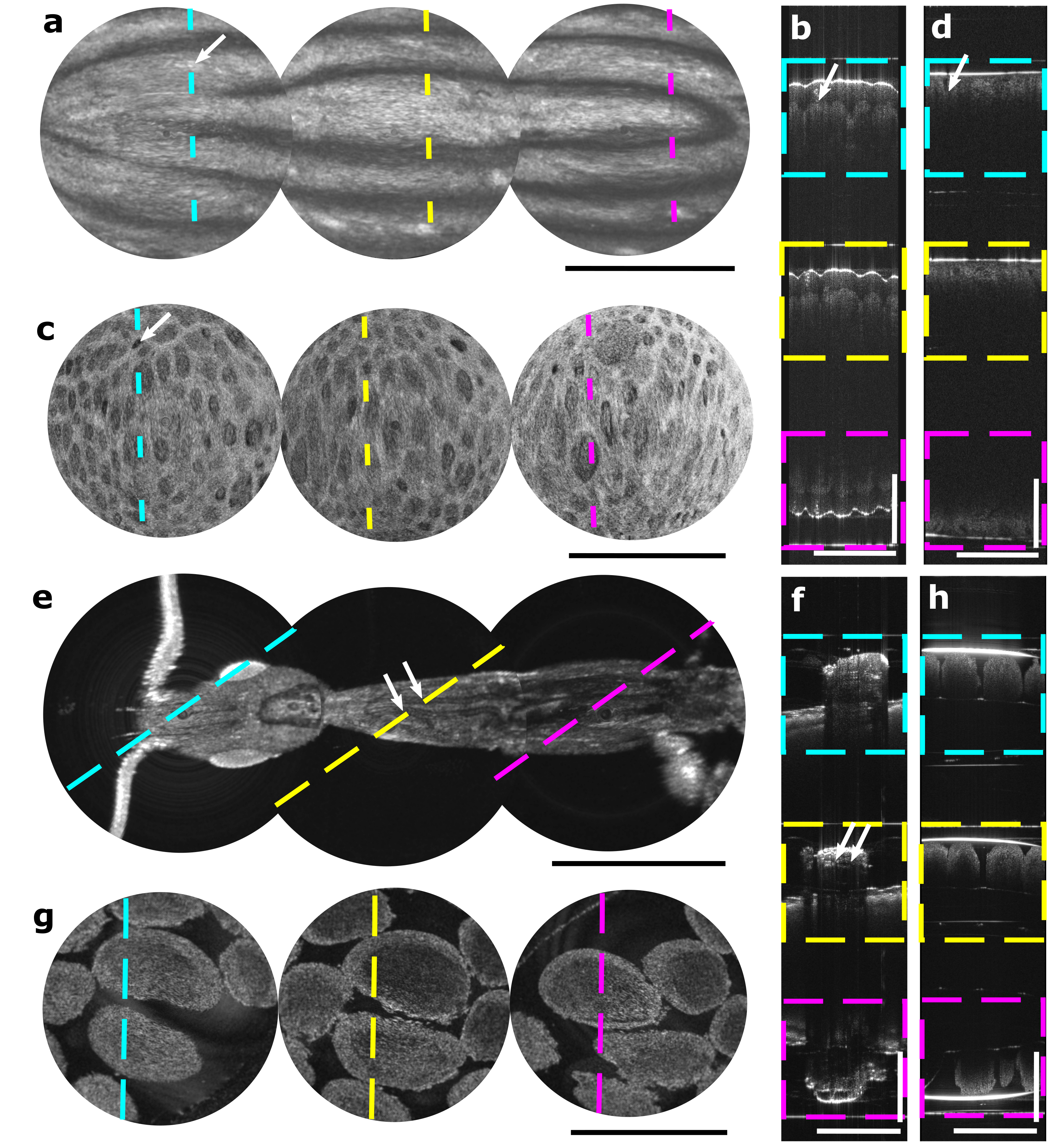}
\caption{OCT imaging of biological samples acquired with spiral scanning. \textit{En face} images of \textbf{(a)} a human finger tip, \textbf{(c)} \textit{ex vivo} pig stomach tissue, \textbf{(e)} an ant and \textbf{(g)} spheroids from three imaging fibers. Cyan, yellow, and magenta lines indicate corresponding depth-multiplexed cross-sectional views \textbf{(b, d, f, h)} from the first, second, and third imaging fiber. Arrows in white indicate a sweat duct in a human finger, a gastric pit in the stomach antrum and the ant's circulatory system and digestive tract. The ant head appeared significantly shorter than the thorax due to its head tilting downward. \textit{En face} images were mean projections of 30 $\mu m$ depth. The bender was actuated at 37.5 V, 500 Hz. Scale bars 1 mm.}
\label{fig:bio}
\end{figure}

Imaging was validated on a wide range of samples including printed targets for calibration and several classes of biological samples. A high-speed swept source OCT system was used as the imaging engine, where the long coherence length of the VCSEL source enabled multiplexing of the fiber array along the depth of an OCT axial scan. In practice, the optical delay of each sample arm length may not appear on an OCT image at the exact designed depth due to manufacturing imperfections. For the third imaging beam, the signal appeared to originate from a depth outside of the designed imaging range due to an imperfect optical delay, resulting in an inverted but otherwise unaffected image (Fig. \ref{fig:bio}b, \ref{fig:bio}d, \ref{fig:bio}f, \ref{fig:bio}h).

\begin{figure}[htbp]
\centering
\includegraphics[width=\linewidth]{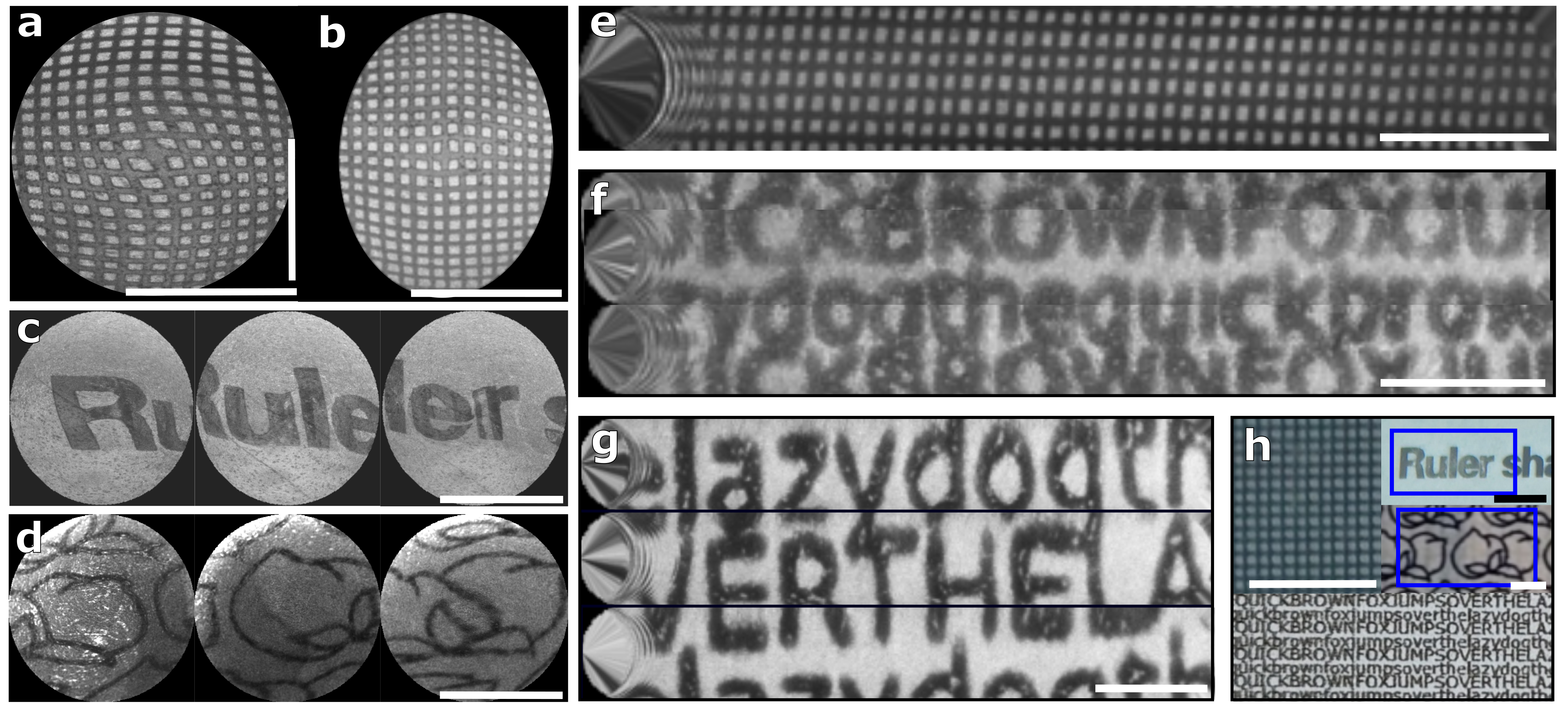}
\caption{OCT imaging of calibration targets acquired with spiral \textbf{(a-d)} and cycloid \textbf{(e-g)} scanning. \textit{En face} images of \textbf{(a, b)} a printed grid of pitch 100 $\mu m$ from spiral scanning, without and with most distortions corrected, \textbf{(c)} a printed word `Ruler' from three imaging fibers, \textbf{(d)} a printed floral pattern on local paper currency from three imaging fibers, \textbf{(e)} a printed grid of pitch 100 $\mu m$ from cycloid scanning, \textbf{(f)} printed sentences (0.5 mm font size) with image fields largely overlapping for continuity, and \textbf{(g)} printed sentences (1 mm font size) with fields non-overlapping and adjacent to maximize coverage. \textbf{(h)} Ground truth photographs of imaging targets which include a printed grid, printed word `Ruler', printed floral pattern on local paper currency, and printed sentences (1 mm font size). \textit{En face} images were mean projections of 30 $\mu m$ depth for images acquired with spiral scanning, and 160 $\mu m$ depth for images acquired with cycloid scanning. The bender was actuated at 37.5 V, 500 Hz. Scale bars 1 mm.}
\label{fig:cali}
\end{figure}

A printed grid was imaged to visualize spiral scan distortions produced by asymmetries in amplitude and phase. Distortions could be largely corrected using a number of simple modifications to the mapping equations (Fig. \ref{fig:cali}a and \ref{fig:cali}b). A larger field of coverage stitching the three FOVs was demonstrated by the imaging of a printed word `Ruler' (Fig. \ref{fig:cali}c) and printed floral patterns on local paper currency (Fig. \ref{fig:cali}d). The FOV from each imaging beam was 2.1  x 1.4  mm, and 2.1 x 2.9  mm when three beams were mosaiced with overlapping regions (Fig. \ref{fig:custom}c and \ref{fig:cali}c). In the cycloid scanning scheme, where the amplitude and phase between the axes of the scan were constant, the image of the printed grid could be fully reconstructed. The FOV of the grid using the cycloid scanning scheme was $\sim$ 0.8 x 5 mm (Fig. \ref{fig:cali}e). A larger field of coverage with large overlapping areas (Fig. \ref{fig:cali}f) and no overlapping areas (Fig. \ref{fig:cali}g) was demonstrated by imaging rows of printed sentences.

The amount of overlap between the FOVs $D_o$ was determined by a combination of design parameters including the fiber spacing $d$, desired FOV size $D_y$ and $D_x$, and final mosaiced image size which was defined based on the application. Non-overlapping fields may be preferred in some biological applications such as maximizing the imaging coverage of adjacent wells on a cell culture plate, while overlapping fields may be preferred for the imaging of larger continuous samples. The overlapping fields facilitated accurate stitching at the cost of some coverage area.

For proof of concept studies in tissue, imaging of a human fingerprint and \textit{ex vivo} pig stomach tissue were performed. The three \textit{en face} views of the fingerprint (Fig. \ref{fig:bio}a) could be approximately stitched to reconstruct a familiar whorl pattern. A sweat duct could be observed in both the \textit{en face} and cross-sectional views of the finger (Fig. \ref{fig:bio}b). \textit{En face} imaging of the stomach (Fig. \ref{fig:bio}c) showed gastric pit architecture, while cross-sections (Fig. \ref{fig:bio}d) showed some columnar structure although the tissue depth penetration was low due to degrading viability of \textit{ex vivo} tissue over time. The capability of visualizing complex tissue structures is important for endoscopic scanning. In principle, a wider FOV from manual stitching of a single areal view swept across a larger zone of interest is possible \cite{lurie_rapid_2015}. However, the limited frame rates or volume rates of advanced modalities, the redundancy of acquiring large numbers of mostly overlapping images, and motion of living tissue make this difficult in practice. Our multi-beam approach enables intrinsic field stitching over an extended area. For larger sweeping coverage, the cycloid scan is an efficient approach that requires only circular `line' scans to build up image areas, with coverage further amplified by multiple beams.

To demonstrate applications in bench-top biology, imaging of a small deceased ant and 3-D spheroids were performed. The subsurface \textit{en face} images as well as cross-sectional images of the ant (Fig. \ref{fig:bio}e and \ref{fig:bio}f) showed deeper internal structures that likely corresponded to vital organs or the digestive tract, providing a unique capability useful for non-destructive functional studies in certain model organisms. The spheroids could also be clearly appreciated in both \textit{en face} and cross-sectional planes (Fig. \ref{fig:bio}g and \ref{fig:bio}h). A wide bender could scan an arbitrary number of optical beams for parallel screening or monitoring applications, such as longitudinal non-destructive monitoring of viability (through the visualization of necrotic core) and treatment response studies on 3-D cultures including more sophisticated organoid models \cite{el-sadek_optical_2020}. 

\section{Discussion}
Piezoelectric tube-actuated resonant fiber scanners have been of great and enduring interest for over two decades, promising microscopic scanning in a tiny package. However, several fundamental limitations of the platform have stymied both technical and translational advances. In this work, we propose a completely new approach to piezoelectric fiber scanning that eliminates the tube and instead uses a powerful and extremely low cost ($<$ US\$10) planar bending actuator to produce a large 2D motion at relatively low voltage. Piezoelectric benders are traditionally understood to be relevant only to linear scanning applications. By harnessing and amplifying the classically undesired coupled vibration in the orthogonal axis, our design is the first to preserve the advantages of a 1D actuator while demonstrating a new functionality as a high performance 2D scanner. The FOV is not limited by the motion of a single fiber but is multiplexed along the bender's width with optical fibers resonating in tandem, uniquely enabled by its planar geometry. The fiber scanning platform is scalable to benders of different sizes and has the potential to be further miniaturized for endoscopic applications. This fresh approach to optical scanning promises multiplexed imaging in a compact package, meeting the high demands of new applications that necessitate high-speed and miniaturized devices. The substantially lower price point at little cost to performance could be an important factor in accelerating commercialization.

2D optical scanning in a small economical package has broad relevance across fields. Spiral scanning would be most applicable in scenarios where a scanner is parked at a specific location of interest. The cycloid scan is a newer approach \cite{liang_cycloid_2018, Yong_high-speed_2010} that can cover much larger areas not limited to the optical aperture size. This approach would be applicable when sweeping a handheld or endoscopic scanner over a large region to generate an image map, or in light-sensitive imaging applications where a continuous laser beam movement is crucial. OCT was the chosen image modality due to the capability of spatial multiplexing by optical path length (OCT depth) using a state-of-the-art long coherence length swept wavelength laser. This enabled each fiber to perform imaging in parallel. High-speed OCT also intrinsically enabled the third and fourth dimension (rapid 3-D) of image acquisition. However, this capability of efficient multi-beam 2D lateral scanning by 1D actuation is highly generalizable to other optical scanning modalities for compact image acquisition or projection designs. For example, each imaging fiber could be multiplexed by time using pulsed illumination and optical path delays if using a single detector, or the fibers could serve separate complementary roles as illumination or detection paths.

This proof of concept study had a number of limitations. In certain scenarios where an endoscope with a shorter rigid tip is required for easier navigation of tight corners, it could be critical to reduce the actuator length. Our optical design had a few disadvantages that could also limit practical use. First, the short working distance of the ball lenses meant there was no room for a front window that would be needed to enclose the scanner in a real-world scenario. Second, the curvature of the imaging plane due to the fiber deflection would likely require further lensing in front of the fibers to achieve a telecentric field. 

Some challenges may present themselves as more practical implementations are pursued. A handheld or endoscopic version of the scanner would need to be relatively insensitive to the orientation of the device relative to gravity, where higher forces on the bender might be required to compensate for its own weight. This could be enabled by replacing 3D printer resin with metallic materials for the clamp, which would better resist flexing and allow stronger screw threading. For further multiplexing with more imaging beams, a very wide bender could in principle accommodate an arbitrarily large number of fibers. In practice, a wider bender would require a larger clamping structure to accommodate a larger bender size and allow larger uniform force to be applied on the bender.
A combination of design parameters such as the clamping structure, clamp forces, and voltage drive should be optimized accordingly for a desired FOV, scan overlap and device footprint. Other important considerations include fiber spacing, which would affect the desired size and shape of the FOV, and fiber resonance frequency, which would be selected partly based on the speed of the imaging system.

\section{Conclusion}
In summary, we presented an innovative approach to miniaturized imaging by the use of a piezoelectric bender for 2D optical scanning. We anticipate a broad range of use cases in sensing and imaging from consumer imaging, medical endoscopy, lidar to neuroscience research, as well as in image projection such as augmented or virtual reality headsets. Important future developments include further improvements to numerical aperture and imaging resolution that would enable other microscopy modalities, and development of a handheld or endoscopic probe with long-term scan repeatability that would justify more demanding deployment applications. 

\section{Acknowledgement}
The authors thank Ko Hui Tan and Ji Yong Lim for laboratory support, and to Dr. Hongwan Liu for valuable discussion.

\section{Funding}
National Research Foundation Singapore (NRFF13-2021-0002); Manufacturing, Trade and Connectivity (MTC) Individual Research Grant (M22K2c0089); Agency for Science, Technology and Research.

\bibliography{ref}

\end{document}